# Uncertainty of data obtained in SRF cavity vertical test

HE Feisi (贺斐思)[1)]
State key laboratory of nuclear physics and technology, Peking University, Beijing, 100871, China

**Abstract:** Vertical test is a commonly used experimental method to qualify Superconducting Radio Frequency (SRF) cavities. Taking the experiences at Jefferson Lab (JLab) in US for example, over thousand of vertical tests have been performed on over 500 different cavities up to now [1]. Most of the tests at JLab followed the method as described in [1], but all the uncertainties of the calculated quality factors as well as the gradients were in-accurate due to the wrong algorithm used. In this paper, a first-principle method was applied to analyze the uncertainty of the data, and the results were compared with those in [1] under typical experiment conditions.

**Key words:** SRF cavity, vertical test, uncertainty
**PACS:** 06.30.Gv, 84.40.-x,

## 1 Introduction and motivation

Over years more than 1000 of vertical tests on over 500 SRF cavities at JLab have been following the algorithm as described in [1]. Though, during the author's visiting at JLab, the algorithm to calculate the uncertainty of intrinsic quality factor (Q0) and gradient (Eacc) in [1] was found to be wrong, since the correlations between variables were not properly treated. In general, it makes the calculated uncertainties in-accurate, e.g. the uncertainty of coupling factor β was under-estimated by more than 25% when it is critically coupled. Another significant consequence was that the uncertainty of Eacc for decay measurement under strongly over-coupled condition was severely over-estimated. In this paper, a first-principle method was applied to the uncertainties of external quality factor of input port (Qe1) and pickup port (Qe2), Q0, and Eacc for decay measurement, as well as to the uncertainty of Q0 and Eacc for continuous wave (CW) measurement.

## 2 Algorithm and assumptions

A first-principle method to calculate random error with any distribution is used:

$$\begin{cases} f = f(x_1, x_2, \dots, x_n). \\ \Delta f = \sqrt{\sum_{i=1}^{n} \left(\frac{\partial f}{\partial x_i}\right)^2 \Delta {x_i}^2}. \end{cases} \quad (1)$$

Note Eq. (1) is valid only if all variables are independent to each other (i.e. uncorrelated). The Δ means standard deviation which is defined as $\Delta y^2 = \langle (y - \langle y \rangle)^2 \rangle$.

In case the power meter readings are well above the noise floor level, all the directly measured data, i.e. decay time constant τ and power meter readings, are uncorrelated. The only figure of merit to judge whether two calculated values are correlated is that whether they both use τ or at least one same power meter reading during the calculation.

Since the uncertainties of measured data are in the form of relative values, it is convenient to rewrite the differential of function f in Eq. (1) as:

$$\frac{df}{f} = \sum_{i=1}^{n} \gamma_i \cdot \frac{dx_i}{x_i}. \quad (2)$$

Applying Eq. (2) to (1), one gets:

$$\frac{\Delta f}{f} = \frac{1}{f}\sqrt{\sum_{i=1}^{n} \left(\gamma_i \frac{f}{x_i}\right)^2 \Delta {x_i}^2} = \sqrt{\sum_{i=1}^{n} \gamma_i^2 \left(\frac{\Delta x_i}{x_i}\right)^2}. \quad (3)$$

So, Qe1, Qe2, Eacc and Q0 will be presented as function of independent variables as in Eq. (2), and the relative error will be calculated with Eq. (3).

Note one more frequently used trick is that:

$$F(x_i) = \prod_j G_j(x_i) \quad \Rightarrow \quad \frac{dF}{F} = \sum_j \frac{dG_j}{G_j}. \quad (4)$$

Eq. (4) is generally valid regardless of the correlations between Gj.

## 3 Define the variable names

In a vertical test, first of all the cables need to be calibrated. Typically the scaling factor Ci, Cr, and Ct need to be determined, which calibrate the power meter readings to the real power at the entrance and exit of the cavity, as shown in Figure 1.

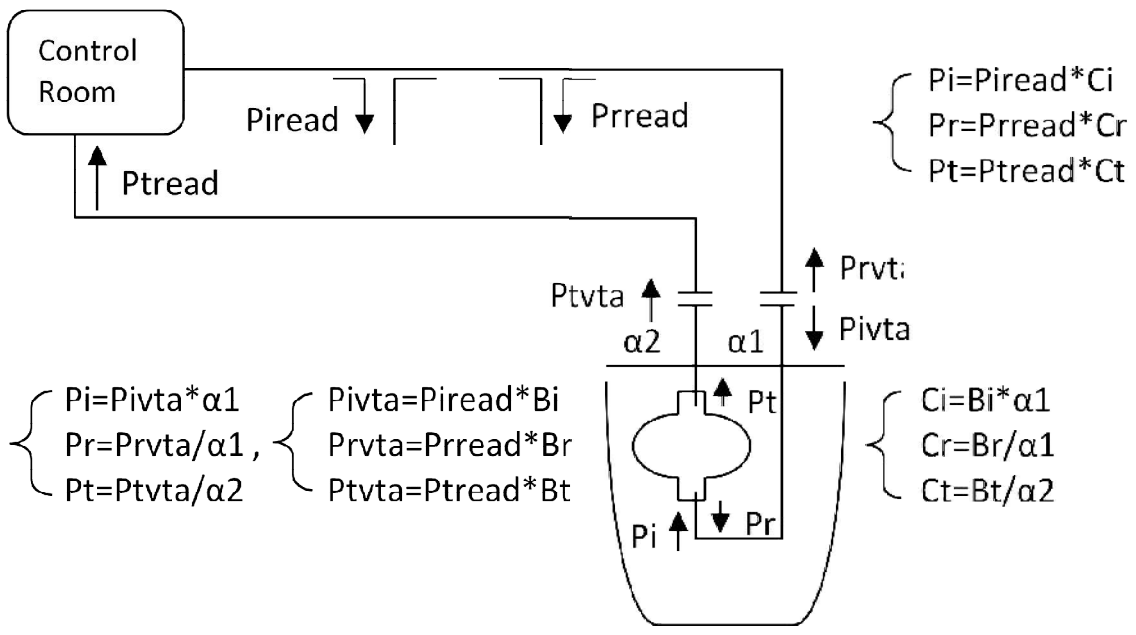

Figure 1: scaling factors for cable calibration

1) Email: heface@pku.edu.cn

| | |
|---|---|
| 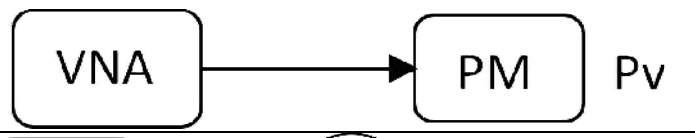  | Pv is the VNA (vector network analyzer) output power from the PM (power meter) reading |
| 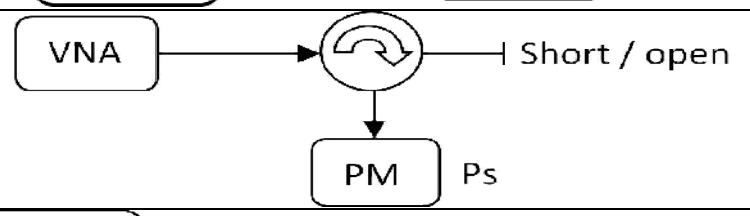  | Ps is the VNA output power with circulator shorted or opened (I usually take the average) |
| 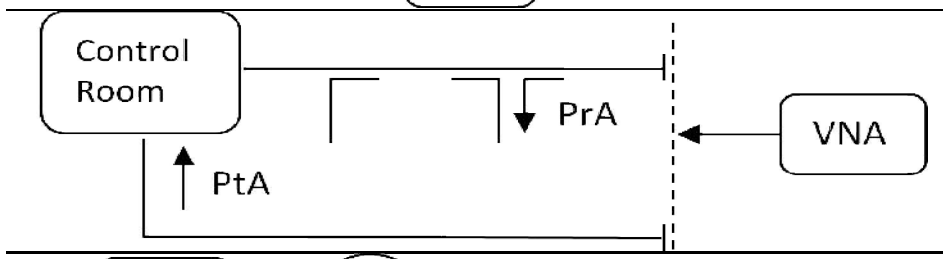  | PrA and PtA are the reflected and transmitted power meter reading in the control room, while using VNA as the RF power source |
| 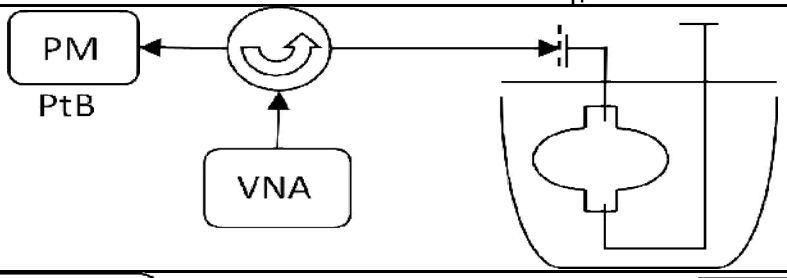  | PtB is the power meter reading when calibrating the damping of transmitted power cable in the dewar. |
| 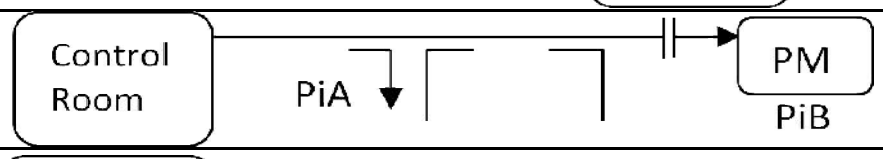  | PiA is the power meter reading in the control room, and PiB is the power meter reading showing real incident power |
| 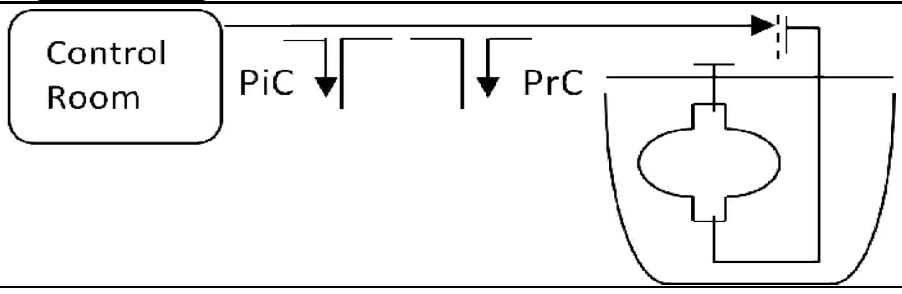  | PiC and PrC are the incident and reflected power meter readings in the control room while the cavity is detuned. |

Figure 2: variable names related to cable calibration

The detailed procedure for cable calibration is well defined in [1]. Here the variable names are defined in Figure 2 for the uncertainty analysis later.

There are two typical measurement techniques: decay measurement which is also used as lower power calibration to determine Qe2; and CW measurement using a known Qe2. For the later one, the Qe2 is obtained by low power measurement; whether the forward power cable damping α1 is re-calibrated at high power level or not will make difference to the calculation of uncertainties. The variable names for Q-E measurement are defined as below:

For decay measurement (which also determine the Qe2), the power meter readings in the control room of incident, reflected, and transmitted power are set as Pir, Prr, and Ptr, respectively. The decay time constant is set as τ.

For high power measurement with known Qe2, the power meter readings in the control room of incident, reflected, and transmitted power are set as PiH, PrH, and PtH, respectively. In case the Qe2 obtained by a lower power measurement is used but the driving cable damping α1 is re-calibrated, then use PiD and PrD to replace the PiC and PrC for the new cable calibration.

## 4 Cable calibration

As defined in the last section, the scaling factors for incident, reflected, and transmitted power are called Ci, Cr, and Ct, respectively. The three scaling factors are correlated, but they could be presented as function of four independent variables as shown in Eq. (5). More details could be found in appendix A.

$$\begin{cases}\theta1 \triangleq \sqrt{\mathrm{Bi/PrA}}, & \theta2 \triangleq \sqrt{\mathrm{PrC/PiC}} \\ \theta3 \triangleq \alpha2/\mathrm{PtA}, & \theta4 \triangleq \sqrt{\mathrm{Pv}}\end{cases} \tag{5}$$
$$\Downarrow$$
$$\mathrm{Ci} = \theta1 \times \theta2 \times \theta4, \mathrm{Cr} = \theta1 \times \theta4/\theta2, \mathrm{Ct} = \theta3 \times \theta4^2.$$

## 5 Decay measurement

There are in all eight independent variables in a decay measurement: θ1, θ2, θ3, θ4, τ, Pir, Prr, and Ptr. The four results obtained from decay measurement, i.e. Qe1, Qe2, Q0, and Eacc, are presented in Eq. (6), and their uncertainties are presented in Eq. (7), (8), (9), and (10). Note the uncertainty of frequency f0 is ignored, and upper sign is for under-coupled. Detailed definition and derivation could be found in appendix B.

$$\begin{cases}\mathrm{Qe1} = \mathrm{Qload} \times \dfrac{(1+\beta^*)}{\beta^*} = \dfrac{4\pi f_0 \tau}{1 \mp |\Gamma|}, & \mathrm{Qe2} = \dfrac{\omega U}{\mathrm{Pt}} = \mathrm{Qload}\dfrac{2\mathrm{Pi}}{\mathrm{Pt}}(1 \mp |\Gamma|) \\ \mathrm{Q0} = \mathrm{Qe2}\dfrac{\mathrm{Pt}}{\mathrm{Pc}} = \mathrm{Qload}\dfrac{2\mathrm{Pi}}{\mathrm{Pi}-\mathrm{Pr}-\mathrm{Pt}}(1 \mp |\Gamma|), & \mathrm{Eacc} = \sqrt{\mathrm{R/Q}}/\mathrm{L} \times \sqrt{\mathrm{Qload} \cdot 2\mathrm{Pi} \cdot (1 \mp |\Gamma|)}\end{cases} \tag{6}$$

$$\frac{\Delta\mathrm{Qe1}}{\mathrm{Qe1}} = \sqrt{\left(\frac{\Delta\tau}{\tau}\right)^2 + \left(\frac{1-\beta1+\beta2}{2\beta1}\right)^2\left(\frac{1}{4}\cdot\left(\frac{\Delta Prr}{Prr}\right)^2 + \frac{1}{4}\cdot\left(\frac{\Delta Pir}{Pir}\right)^2 + \left(\frac{\Delta\theta2}{\theta2}\right)^2\right)} \tag{7}$$

$$\frac{\Delta\mathrm{Qe2}}{\mathrm{Qe2}} = \sqrt{\begin{array}{l}\left(\frac{\Delta\tau}{\tau}\right)^2 + \left(\frac{\Delta\theta1}{\theta1}\right)^2 + \left(\frac{1+\beta1+\beta2}{2\beta1}\right)^2\left(\frac{\Delta\theta2}{\theta2}\right)^2 + \left(\frac{\Delta\theta3}{\theta3}\right)^2 + \left(\frac{\Delta\theta4}{\theta4}\right)^2 \\ + \left(\frac{\Delta Ptr}{Ptr}\right)^2 + \left(\frac{1+3\beta1+\beta2}{4\beta1}\right)^2\left(\frac{\Delta Pir}{Pir}\right)^2 + \left(\frac{1-\beta1+\beta2}{4\beta1}\right)^2\left(\frac{\Delta Prr}{Prr}\right)^2\end{array}} \tag{8}$$

) Email: heface@pku.edu.cn

$$\frac{\Delta Q0}{Q0}=\sqrt{\begin{aligned}&\left(\frac{\Delta\tau}{\tau}\right)^2+\beta2^2\left(\frac{\Delta\theta1}{\theta1}\right)^2+\left(\frac{\beta1-\beta1^2-\beta2-\beta2^2}{2\beta1}\right)^2\cdot\left(\frac{\Delta\theta2}{\theta2}\right)^2+\beta2^2\cdot\left[\left(\frac{\Delta\theta3}{\theta3}\right)^2+\left(\frac{\Delta\theta4}{\theta4}\right)^2+\left(\frac{\Delta Ptr}{Ptr}\right)^2\right]\\&+\left(\frac{(\beta1-\beta2)(1-\beta1+\beta2)}{4\beta1}\right)^2\left(\frac{\Delta Prr}{Prr}\right)^2+\left(\frac{\beta1-\beta1^2-\beta2-\beta2^2-2\beta1\beta2}{4\beta1}\right)^2\left(\frac{\Delta Pir}{Pir}\right)^2\end{aligned}} \tag{9}$$

$$\frac{\Delta Eacc}{Eacc}=\frac{1}{2}\sqrt{\begin{aligned}&\left(\frac{\Delta\tau}{\tau}\right)^2+\left(\frac{\Delta\theta1}{\theta1}\right)^2+\left(\frac{1+\beta1+\beta2}{2\beta1}\right)^2\left(\frac{\Delta\theta2}{\theta2}\right)^2+\left(\frac{\Delta\theta4}{\theta4}\right)^2\\&+\left(\frac{1+3\beta1+\beta2}{4\beta1}\right)^2\left(\frac{\Delta Pir}{Pir}\right)^2+\left(\frac{1-\beta1+\beta2}{4\beta1}\right)^2\left(\frac{\Delta Prr}{Prr}\right)^2\end{aligned}} \tag{10}$$

## 6 CW measurement with known Qe2

If the cable calibrations are not changed after the decay measurement which determines the Qe2, then there are in all eleven independent variables, i.e. eight same as in the decay measurement, and PiH, PrH and PtH from the high power measurement readings. The uncertainty of Q0 and Eacc are shown in Eq. (11) and (13)

But in case the cable loss on the RF driving cable is changed by heating effect, the cable loss factor α1 will usually be re-calibrated. In this case, there are two more independent variables PiD and PrD as defined in section 3. Accordingly, define $\theta5=\sqrt{PrD/PiD}$. The uncertainty of Q0 is different, as shown in Eq. (12), and Eacc is the same. Detailed definitions and derivation could be found in appendix D.

$$\frac{\Delta Q0}{Q0}=\sqrt{\begin{aligned}&\left(\frac{\Delta\tau}{\tau}\right)^2+\beta2cw^2\left(\frac{\Delta\theta1}{\theta1}\right)^2+\left(\frac{\beta1cw+\beta1cw\beta2-\beta1((-1+\beta1cw)\beta1cw+(1+\beta2cw)^2)}{2\beta1\beta1cw}\right)^2\cdot\left(\frac{\Delta\theta2}{\theta2}\right)^2\\&+\beta2cw^2\cdot\left[\left(\frac{\Delta\theta3}{\theta3}\right)^2+\left(\frac{\Delta\theta4}{\theta4}\right)^2\right]+\left(\frac{\Delta Ptr}{Ptr}\right)^2+(1+\beta2cw)^2\left(\frac{\Delta PtH}{PtH}\right)^2+\frac{(1+\beta1cw+\beta2cw)^4}{16\cdot\beta1cw^2}\left(\frac{\Delta PiH}{PiH}\right)^2\\&+\left(\frac{1+3\beta1+\beta2}{4\beta1}\right)^2\left(\frac{\Delta Pir}{Pir}\right)^2+\frac{(1-\beta1cw+\beta2cw)^4}{16\cdot\beta1cw^2}\left(\frac{\Delta PrH}{PrH}\right)^2+\left(\frac{1-\beta1+\beta2}{4\beta1}\right)^2\left(\frac{\Delta Prr}{Prr}\right)^2\end{aligned}} \tag{11}$$

$$\left.\frac{\Delta Q0}{Q0}\right|_{re-cal}=\sqrt{\begin{aligned}&\left(\frac{\Delta\tau}{\tau}\right)^2+\beta2cw^2\left(\frac{\Delta\theta1}{\theta1}\right)^2+\left(\frac{1+\beta1+\beta2}{2\beta1}\right)^2\cdot\left(\frac{\Delta\theta2}{\theta2}\right)^2+\left(\frac{\beta1cw^2+(1+\beta2cw)^2}{2\beta1cw}\right)^2\cdot\left(\frac{\Delta\theta5}{\theta5}\right)^2\\&+\beta2cw^2\cdot\left[\left(\frac{\Delta\theta3}{\theta3}\right)^2+\left(\frac{\Delta\theta4}{\theta4}\right)^2\right]+\left(\frac{\Delta Ptr}{Ptr}\right)^2+(1+\beta2cw)^2\left(\frac{\Delta PtH}{PtH}\right)^2+\frac{(1+\beta1cw+\beta2cw)^4}{16\cdot\beta1cw^2}\left(\frac{\Delta PiH}{PiH}\right)^2\\&+\left(\frac{1+3\beta1+\beta2}{4\beta1}\right)^2\left(\frac{\Delta Pir}{Pir}\right)^2+\frac{(1-\beta1cw+\beta2cw)^4}{16\cdot\beta1cw^2}\left(\frac{\Delta PrH}{PrH}\right)^2+\left(\frac{1-\beta1+\beta2}{4\beta1}\right)^2\left(\frac{\Delta Prr}{Prr}\right)^2\end{aligned}} \tag{12}$$

$$\frac{\Delta Eacc}{Eacc}=\frac{1}{2}\sqrt{\begin{aligned}&\left(\frac{\Delta\tau}{\tau}\right)^2+\left(\frac{\Delta\theta1}{\theta1}\right)^2+\left(\frac{1+\beta1+\beta2}{2\beta1}\right)^2\left(\frac{\Delta\theta2}{\theta2}\right)^2+\left(\frac{\Delta\theta4}{\theta4}\right)^2+\left(\frac{\Delta Ptr}{Ptr}\right)^2\\&+\left(\frac{1+3\beta1+\beta2}{4\beta1}\right)^2\left(\frac{\Delta Pir}{Pir}\right)^2+\left(\frac{1-\beta1+\beta2}{4\beta1}\right)^2\left(\frac{\Delta Prr}{Prr}\right)^2+\left(\frac{\Delta PtH}{PtH}\right)^2\end{aligned}} \tag{13}$$

## 7 Calculated typical uncertainties

The uncertainty of Q0 and Eacc for decay measurement are obtained in Eq. (9) and (10), while for CW measurement with known Qe2 they are obtained in Eq. (11), (12), and(13).

For a typical vertical test, the tolerance of power meter readings in control room during the test (i.e. Pir, Prr, Ptr, PiH, PrH, PtH) is within 5% when attenuators are well distributed in the low level control system. For cable calibration, uncertainty of Pv, PtA, and PiB are 5% too. Additional error induced by the standing wave in circulator makes the tolerance of Ps and PtB 6-7%. The standing wave introduces about 7% extra error to the sampling of directional coupler, which makes a tolerance of about 10% to PrA, PrC, PrD, PiA, PiC, and PiD.

The trend of calculated uncertainties are illustrated in Figure 3, following the assumptions above, and assuming β2 is 0.02, β1 is 2.5 when calibrating Qe2 for CW measurement, and accuracy of decay time constant is 3%.

Note a significant source of uncertainty when it is way off critical coupling is that the difference between Pi and Pr have big error bar. Though, from Eq. (6) it is noticed that uncertainties of Q0 and Eac should not diverge when it is under-coupled and over-coupled, respectively. It agrees very well with the curve shown in Figure 3.

Note: in case the coefficient Bi, Br, Bt, and α2 are measured at multiple power levels, and the averages are taken for each of them, then the correlations between them become much weaker, and it is reasonable to treat Bi, Br, α1, and Bt/ α2 as four independent variables instead of θ1-θ4. Eq. (9)-(13) could be re-derived accordingly.

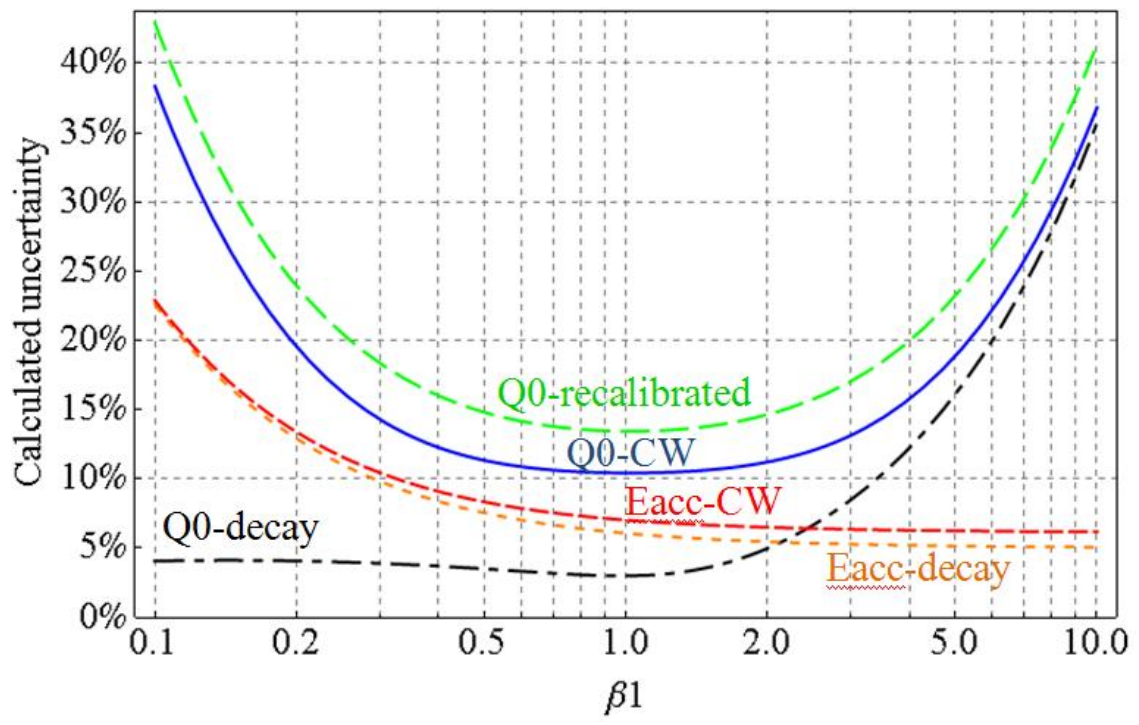

Figure 3: trend of calculated error vs. input coupling

If the wrong algorithm in [1] is used, which doesn't consider the correlations in between variables, then the differences of Q0 and Eacc for decay measurement are illustrated in Figure 4. Note same assumptions are used as in Figure 3, together with the fact that tolerance of 5% is typically assumed for all the power meter readings when using algorithm in [1]. Note detailed calculation could be found in [1], and it is recalled in appendix D.

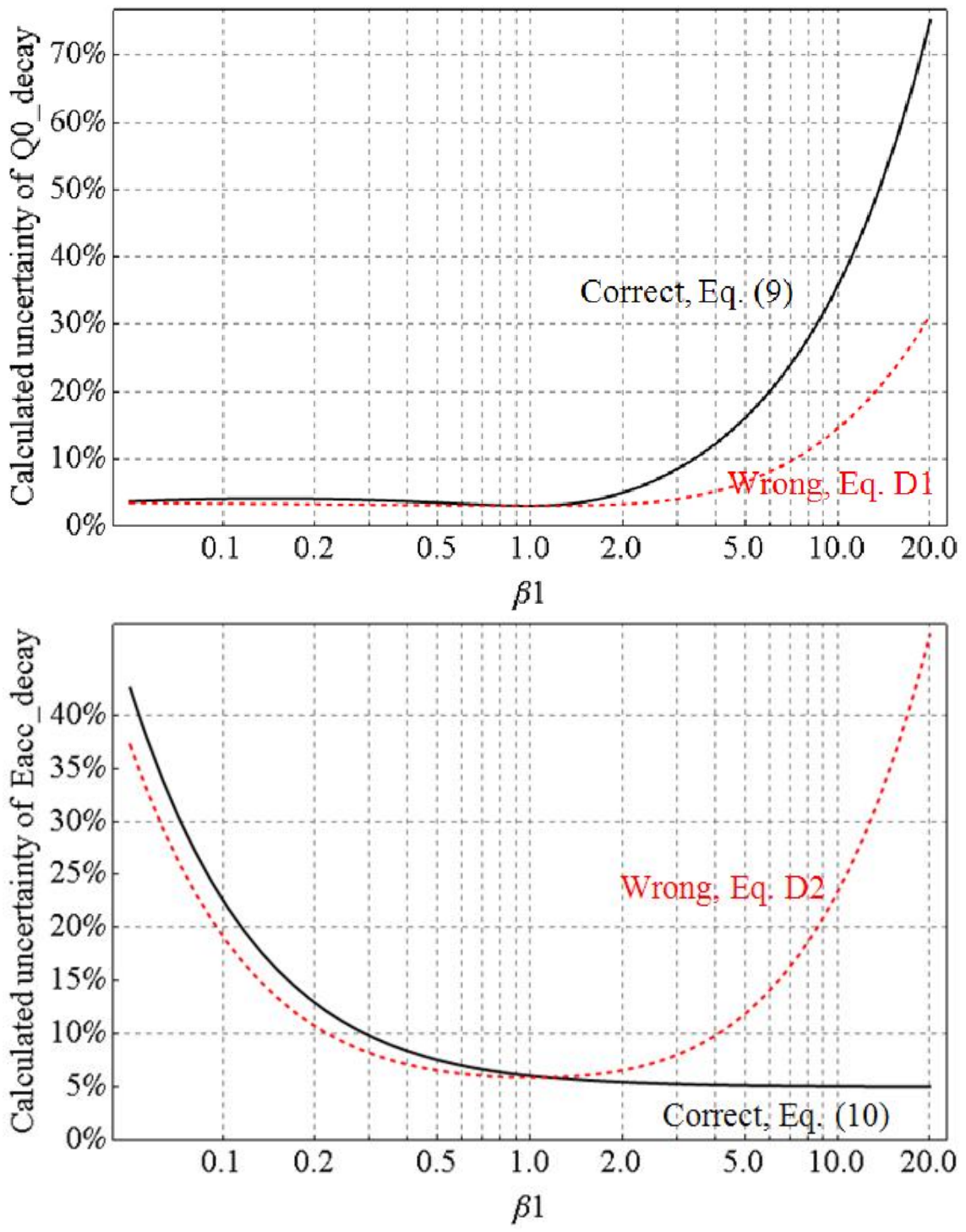

Figure 4: Comparison of calculated error between the correct and wrong algorithm

## 8 Conclusion

The wrong uncertainty calculation method for processing SRF cavity vertical test data at JLab is corrected in this paper. The formula of uncertainty calculation for decay and CW measurement is provided in Eq. (9) to (13). The trend of uncertainty is illustrated as the input coupling changes, and the accuracy with decay measurement is found better than CW measurement. The suggestion in [1] that CW measurement should be performed with 0.5<β<2 is still valid.

## 9 Acknowledgement

I would like to appreciate Tom Powers at JLab for providing the information on how standing wave affects the accuracy of power meter readings. I would also like to appreciate Gianluigi Ciovati at JLab for the discussion on how averaging affects the calculation. Moreover, I would like to appreciate Haipeng Wang at JLab for the suggestions on making this paper more readable.

## Appendix A: present cable calibration coefficients as function of independent variables

During the cable calibration, there are in all nine independent power meter readings: Pv, Ps, PrA, PtA, PiA, PiB, PtB, PiC, and PrC. Usually Bi, Br, Bt, α1, α2, Ci, Cr, and Ct are calculated from those measured values as below:

$$\begin{cases} \mathrm{Br} = \mathrm{Pv}/\mathrm{PrA}, \quad \mathrm{Bt} = \mathrm{Pv}/\mathrm{PtA}, \quad \mathrm{Bi} = \mathrm{PiB}/\mathrm{PiA} \\ \alpha1 = \sqrt{\dfrac{\mathrm{PrC}\times\mathrm{Br}}{\mathrm{PiC}\times\mathrm{Bi}}}, \quad \alpha2 = \sqrt{\mathrm{PtB}/\mathrm{Ps}} \\ \mathrm{Ci} = \mathrm{Bi}\cdot\alpha1 = \sqrt{\dfrac{\mathrm{PrC}}{\mathrm{PiC}}}\times\sqrt{\dfrac{\mathrm{Bi}}{\mathrm{PrA}}}\times\sqrt{\mathrm{Pv}} \\ \mathrm{Cr} = \dfrac{\mathrm{Br}}{\alpha1} = \left(\sqrt{\dfrac{\mathrm{PrC}}{\mathrm{PiC}}}\right)^{-1}\times\sqrt{\dfrac{\mathrm{Bi}}{\mathrm{PrA}}}\times\sqrt{\mathrm{Pv}} \\ \mathrm{Ct} = \dfrac{\mathrm{Bt}}{\alpha2} = \dfrac{\alpha2}{\mathrm{PtA}}\times\mathrm{Pv} \end{cases} \tag{A1}$$

Apparently Ci, Cr, and Ct are correlated. But they could be presented as functions of four independent variables defined as θ1, θ2, θ3 and θ4 in Eq. (5), which is recalled as below.

$$\begin{cases} \theta1 \triangleq \sqrt{\mathrm{Bi}/\mathrm{PrA}}, \quad \theta2 \triangleq \sqrt{\mathrm{PrC}/\mathrm{PiC}} \\ \theta3 \triangleq \alpha2/\mathrm{PtA}, \quad \theta4 \triangleq \sqrt{\mathrm{Pv}} \end{cases}$$

$$\Downarrow$$

$$\mathrm{Ci} = \theta1\times\theta2\times\theta4, \mathrm{Cr} = \theta1\times\theta4/\theta2, \mathrm{Ct} = \theta3\times\theta4^2.$$

The uncertainties of θ1 to θ4 are calculated in Eq.(A2)

$$\begin{cases} \dfrac{\Delta\theta1}{\theta1} = \dfrac{1}{2}\sqrt{\left(\dfrac{\Delta\mathrm{PiB}}{\mathrm{PiB}}\right)^2 + \left(\dfrac{\Delta\mathrm{PiA}}{\mathrm{PiA}}\right)^2 + \left(\dfrac{\Delta\mathrm{PrA}}{\mathrm{PrA}}\right)^2} \\ \dfrac{\Delta\theta2}{\theta2} = \dfrac{1}{2}\sqrt{\left(\dfrac{\Delta\mathrm{PrC}}{\mathrm{PrC}}\right)^2 + \left(\dfrac{\Delta\mathrm{PiC}}{\mathrm{PiC}}\right)^2} \\ \dfrac{\Delta\theta3}{\theta3} = \sqrt{\dfrac{1}{4}\left(\dfrac{\Delta\mathrm{PtB}}{\mathrm{PtB}}\right)^2 + \dfrac{1}{4}\left(\dfrac{\Delta\mathrm{Ps}}{\mathrm{Ps}}\right)^2 + \left(\dfrac{\Delta\mathrm{PtA}}{\mathrm{PtA}}\right)^2} \\ \dfrac{\Delta\theta4}{\theta4} = \dfrac{1}{2}\left(\dfrac{\Delta\mathrm{Pv}}{\mathrm{Pv}}\right) \end{cases} \tag{A2}$$

## Appendix B: uncertainty analysis for decay measurement

As defined in section 3, there are in all eight independent variables in a decay measurement: θ1, θ2, θ3, θ4, τ, Pir, Prr, and Ptr. Some variables that help the calculation are listed as below:

$$
\begin{cases}
\mathrm{Pi} = \mathrm{Ci}\cdot\mathrm{Pir} = \theta1\cdot\theta2\cdot\theta4\cdot\mathrm{Pir}, \quad \mathrm{Pr} = \mathrm{Cr}\cdot\mathrm{Prr} = \theta1\cdot\dfrac{\theta4}{\theta2}\cdot\mathrm{Pir}, \quad \mathrm{Pt} = \mathrm{Ct}\cdot\mathrm{Ptr} = \theta3\cdot\theta4^2\cdot\mathrm{Ptr} \\
\text{Reflection } |\Gamma| = \sqrt{\dfrac{\mathrm{Pr}}{\mathrm{Pi}}} = \dfrac{1}{\theta2}\sqrt{\dfrac{\mathrm{Prr}}{\mathrm{Pir}}}, \quad \beta^* = \dfrac{Pe}{Pi - Pr} = \dfrac{1 \mp |\Gamma|}{1 \pm |\Gamma|}, \textbf{where upper is for under coupled} \\
\text{Wall loss } \mathrm{Pc} = \mathrm{Pi} - \mathrm{Pr} - \mathrm{Pt}, \quad \mathrm{Qload} = \dfrac{\omega U}{\mathrm{Pe} + \mathrm{Pc} + \mathrm{Pt}} = 2\pi f_0 \times \tau \\
\beta2 = \mathrm{Pt}/\mathrm{Pc}, \quad \beta1 = \mathrm{Pe}/\mathrm{Pc} = \beta^*(1 + \beta2) \\
f1 \triangleq \dfrac{\pm|\Gamma|}{(1 \pm |\Gamma|)} = \pm\dfrac{\dfrac{1}{\theta2}\sqrt{\dfrac{\mathrm{Prr}}{\mathrm{Pir}}}}{\left(1 \pm \dfrac{1}{\theta2}\sqrt{\dfrac{\mathrm{Prr}}{\mathrm{Pir}}}\right)} = \dfrac{1 - \beta1 + \beta2}{2(1 + \beta2)} \\
f2 \triangleq \dfrac{\mathrm{Pi}}{\mathrm{Pi} - \mathrm{Pr}} = \dfrac{\theta2\cdot\mathrm{Pir}}{(\theta2\cdot\mathrm{Pir} - \mathrm{Prr}/\theta2)} = \dfrac{(1 + \beta1 + \beta2)^2}{4\cdot\beta1(1 + \beta2)} \\
f3 \triangleq \dfrac{\mathrm{Pr}}{\mathrm{Pi} - \mathrm{Pr}} = \dfrac{\mathrm{Prr}/\theta2}{(\theta2\cdot\mathrm{Pir} - \mathrm{Prr}/\theta2)} = \dfrac{(1 - \beta1 + \beta2)^2}{4\cdot\beta1(1 + \beta2)} \\
f4 \triangleq \dfrac{\mathrm{Pi}}{\mathrm{Pc}} = \dfrac{\theta1\cdot\theta2\cdot\mathrm{Pir}}{\theta1(\theta2\mathrm{Pir} - \mathrm{Prr}/\theta2) - \theta3\theta4\mathrm{Ptr}} = \dfrac{(1 + \beta1 + \beta2)^2}{4\beta1} \\
f5 \triangleq \dfrac{\mathrm{Pr}}{\mathrm{Pc}} = \dfrac{\theta1/\theta2\cdot\mathrm{Prr}}{\theta1(\theta2\mathrm{Pir} - \mathrm{Prr}/\theta2) - \theta3\theta4\mathrm{Ptr}} = \dfrac{(1 - \beta1 + \beta2)^2}{4\beta1} \\
f6 \triangleq \dfrac{\mathrm{Pt}}{\mathrm{Pc}} = \dfrac{\theta3\cdot\theta4\cdot\mathrm{Ptr}}{\theta1(\theta2\mathrm{Pir} - \mathrm{Prr}/\theta2) - \theta3\theta4\mathrm{Ptr}} = \beta_2
\end{cases}
$$

Recall Eq. (6) for the calculation of final results as below:

$$
\begin{cases}
\mathrm{Qe1} = \dfrac{\omega U}{\mathrm{Pe}} = \mathrm{Qload}\times\dfrac{(1 + \beta^*)}{\beta^*} = \dfrac{4\pi f_0\tau}{1 \mp |\Gamma|}, \quad \mathrm{Qe2} = \dfrac{\omega U}{\mathrm{Pt}} = \mathrm{Qload}(1 + \beta^*)\dfrac{\mathrm{Pi} - \mathrm{Pr}}{\mathrm{Pt}} = \mathrm{Qload}\dfrac{2\mathrm{Pi}}{\mathrm{Pt}}(1 \mp |\Gamma|) \\
\mathrm{Q0} = \mathrm{Qe2}\dfrac{\mathrm{Pt}}{\mathrm{Pc}} = \mathrm{Qload}\dfrac{2\mathrm{Pi}}{\mathrm{Pi} - \mathrm{Pr} - \mathrm{Pt}}(1 \mp |\Gamma|), \quad \mathrm{Eacc} = \dfrac{\sqrt{\dfrac{R}{Q}}}{L}\times\sqrt{\mathrm{Pt}\times\mathrm{Qe2}} = \dfrac{\sqrt{\dfrac{R}{Q}}}{L}\times\sqrt{\mathrm{Qload}\cdot 2\mathrm{Pi}\cdot(1 \mp |\Gamma|)}
\end{cases}
$$

The error on Qe1 is derived as below. It gives the Eq. (7):

$$
\frac{d\mathrm{Qe1}}{\mathrm{Qe1}} = \frac{d\tau}{\tau} \pm \frac{d|\Gamma|}{1 \mp |\Gamma|} = \frac{d\tau}{\tau} \pm \frac{|\Gamma|}{1 \mp |\Gamma|}\left(\frac{1}{2}\cdot\frac{dPrr}{Prr} - \frac{1}{2}\cdot\frac{dPir}{Pir} - \frac{d\theta2}{\theta2}\right)
$$

$$
\frac{\Delta\mathrm{Qe1}}{\mathrm{Qe1}} = \sqrt{\left(\frac{\Delta\tau}{\tau}\right)^2 + \left(\frac{1 - \beta1 + \beta2}{2\beta1}\right)^2\left(\frac{1}{4}\cdot\left(\frac{\Delta Prr}{Prr}\right)^2 + \frac{1}{4}\cdot\left(\frac{\Delta Pir}{Pir}\right)^2 + \left(\frac{\Delta\theta2}{\theta2}\right)^2\right)}
$$

The error on Qe2 is derived as below. It gives the Eq. (8):

$$
\begin{aligned}
\frac{d\mathrm{Qe2}}{\mathrm{Qe2}} &= \frac{d\tau}{\tau} + \frac{d\theta1}{\theta1} + (f1 + f2 + f3)\frac{d\theta2}{\theta2} - \frac{d\theta3}{\theta3} - \frac{d\theta4}{\theta4} - \frac{dPtr}{Ptr} + \left(\frac{f1}{2} + f2\right)\frac{dPir}{Pir} - \left(\frac{f1}{2} + f3\right)\frac{dPrr}{Prr} \\
&= \frac{d\tau}{\tau} + \frac{d\theta1}{\theta1} + \frac{1 + \beta1 + \beta2}{2\beta1}\frac{d\theta2}{\theta2} - \frac{d\theta3}{\theta3} - \frac{d\theta4}{\theta4} - \frac{dPtr}{Ptr} + \frac{1 + 3\beta1 + \beta2}{4\beta1}\frac{dPir}{Pir} - \frac{1 - \beta1 + \beta2}{4\beta1}\frac{dPrr}{Prr}
\end{aligned}
$$

$$
\frac{\Delta\mathrm{Qe2}}{\mathrm{Qe2}} = \sqrt{\begin{gathered}\left(\frac{\Delta\tau}{\tau}\right)^2 + \left(\frac{\Delta\theta1}{\theta1}\right)^2 + \left(\frac{1 + \beta1 + \beta2}{2\beta1}\right)^2\left(\frac{\Delta\theta2}{\theta2}\right)^2 + \left(\frac{\Delta\theta3}{\theta3}\right)^2 + \left(\frac{\Delta\theta4}{\theta4}\right)^2 \\ + \left(\frac{\Delta Ptr}{Ptr}\right)^2 + \left(\frac{1 + 3\beta1 + \beta2}{4\beta1}\right)^2\left(\frac{\Delta Pir}{Pir}\right)^2 + \left(\frac{1 - \beta1 + \beta2}{4\beta1}\right)^2\left(\frac{\Delta Prr}{Prr}\right)^2\end{gathered}}
$$

The error on Q0 is derived as below. It gives the Eq. (9):

$$
\begin{aligned}
\frac{d\mathrm{Q0}}{\mathrm{Q0}} &= \frac{d\tau}{\tau} + (1 - f4 + f5)\cdot\frac{d\theta1}{\theta1} + (f1 + f2 + f3 - f4 - f5)\frac{d\theta2}{\theta2} + f6\cdot\left(\frac{d\theta3}{\theta3} + \frac{d\theta4}{\theta4} + \frac{dPtr}{Ptr}\right) \\
&\quad + \left(\frac{f1}{2} + f2 - f4\right)\frac{dPir}{Pir} - \left(\frac{f1}{2} + f3 - f5\right)\frac{dPrr}{Prr} \\
&= \frac{d\tau}{\tau} - \beta2\cdot\frac{d\theta1}{\theta1} + \frac{\beta1 - \beta1^2 - \beta2 - \beta2^2}{2\beta1}\frac{d\theta2}{\theta2} + \beta2\cdot\left(\frac{d\theta3}{\theta3} + \frac{d\theta4}{\theta4} + \frac{dPtr}{Ptr}\right) \\
&\quad + \frac{\beta1 - \beta1^2 - \beta2 - \beta2^2 - 2\beta1\beta2}{4\beta1}\frac{dPir}{Pir} - \frac{(\beta1 - \beta2)(1 - \beta1 + \beta2)}{4\beta1}\frac{dPrr}{Prr}
\end{aligned}
$$

$$\frac{\Delta Q0}{Q0}=\sqrt{\left(\frac{\Delta\tau}{\tau}\right)^2+\beta 2^2\left(\frac{\Delta\theta 1}{\theta 1}\right)^2+\left(\frac{\beta 1-\beta 1^2-\beta 2-\beta 2^2}{2\beta 1}\right)^2\cdot\left(\frac{\Delta\theta 2}{\theta 2}\right)^2+\beta 2^2\cdot\left[\left(\frac{\Delta\theta 3}{\theta 3}\right)^2+\left(\frac{\Delta\theta 4}{\theta 4}\right)^2+\left(\frac{\Delta Ptr}{Ptr}\right)^2\right]+\left(\frac{(\beta 1-\beta 2)(1-\beta 1+\beta 2)}{4\beta 1}\right)^2\left(\frac{\Delta Prr}{Prr}\right)^2+\left(\frac{\beta 1-\beta 1^2-\beta 2-\beta 2^2-2\beta 1\beta 2}{4\beta 1}\right)^2\left(\frac{\Delta Pir}{Pir}\right)^2}$$

The error on Eacc is derived as below. It gives the Eq. (10):

$$\begin{aligned}\frac{dEacc}{Eacc}&=\frac{1}{2}\frac{d\tau}{\tau}+\frac{1}{2}\frac{d\theta 1}{\theta 1}+(f1+f2+f3)\frac{d\theta 2}{\theta 2}+\frac{1}{2}\frac{d\theta 4}{\theta 4}+\left(\frac{f1}{2}+f2\right)\frac{dPir}{Pir}-\left(\frac{f1}{2}+f3\right)\frac{dPrr}{Prr}\\&=\frac{1}{2}\frac{d\tau}{\tau}+\frac{1}{2}\frac{d\theta 1}{\theta 1}+\frac{1+\beta 1+\beta 2}{4\beta 1}\frac{d\theta 2}{\theta 2}+\frac{1}{2}\frac{d\theta 4}{\theta 4}+\frac{1+3\beta 1+\beta 2}{8\beta 1}\frac{dPir}{Pir}-\frac{1-\beta 1+\beta 2}{8\beta 1}\frac{dPrr}{Prr}\end{aligned}$$

$$\frac{\Delta Eacc}{Eacc}=\frac{1}{2}\sqrt{\left(\frac{\Delta\tau}{\tau}\right)^2+\left(\frac{\Delta\theta 1}{\theta 1}\right)^2+\left(\frac{1+\beta 1+\beta 2}{2\beta 1}\right)^2\left(\frac{\Delta\theta 2}{\theta 2}\right)^2+\left(\frac{\Delta\theta 4}{\theta 4}\right)^2+\left(\frac{1+3\beta 1+\beta 2}{4\beta 1}\right)^2\left(\frac{\Delta Pir}{Pir}\right)^2+\left(\frac{1-\beta 1+\beta 2}{4\beta 1}\right)^2\left(\frac{\Delta Prr}{Prr}\right)^2}$$

## Appendix C: uncertainty analysis for CW measurement with known Qe2

### C.1: the cable calibrations are not changed after the decay measurement which determines the Qe2.

In this case there are in all eleven independent variables, i.e. eight same as in the decay measurement, and PiH, PrH and PtH from the high power measurement readings. The Q0 and Eacc are calculated using Eq. (C1).

$$\begin{cases}Q0=Qe2\times\dfrac{\theta 3\cdot\theta 4\cdot PtH}{\theta 1(\theta 2PiH-PrH/\theta 2)-\theta 3\theta 4PtH}\\ Eacc=\dfrac{\sqrt{R/Q}}{L}\times\sqrt{Qe2\times\theta 3\cdot\theta 4^2\cdot PtH}\end{cases}\qquad(C1)$$

Some more variables are needed to help the calculation. Set β1cw and β2cw similar to the decay measurement, with Pir, Prr and Ptr replaced by PiH, PrH, and PtH, respectively. Then define f7, f8 and f9 as following. Note definition of f1, f2, and f3 could be found in Appendix C.

$$f7\triangleq\frac{Picw}{Pccw}=\frac{\theta 1\cdot\theta 2\cdot PiH}{\theta 1(\theta 2PiH-PrH/\theta 2)-\theta 3\theta 4PtH}=\frac{(1+\beta 1cw+\beta 2cw)^2}{4\beta 1cw}$$

$$f8\triangleq\frac{Prcw}{Pccw}=\frac{\theta 1/\theta 2\cdot PrH}{\theta 1(\theta 2PiH-PrH/\theta 2)-\theta 3\theta 4PtH}=\frac{(1-\beta 1cw+\beta 2cw)^2}{4\beta 1cw}$$

$$f9\triangleq\frac{Ptcw}{Pccw}=\frac{\theta 3\cdot\theta 4\cdot PtH}{\theta 1(\theta 2PiH-PrH/\theta 2)-\theta 3\theta 4PtH}=\beta 2cw$$

The error on Q0 is derived as below. It gives the Eq. (11)

$$\begin{aligned}\frac{dQ0}{Q0}&=\frac{d\tau}{\tau}+(1-f7+f8)\cdot\frac{d\theta 1}{\theta 1}+(f1+f2+f3-f7-f8)\frac{d\theta 2}{\theta 2}+f9\cdot\left(\frac{d\theta 3}{\theta 3}+\frac{d\theta 4}{\theta 4}\right)\\&-\frac{dPtr}{Ptr}+(1+f9)\frac{dPtH}{PtH}+\left(\frac{f1}{2}+f2\right)\frac{dPir}{Pir}-f7\cdot\frac{dPiH}{PiH}-\left(\frac{f1}{2}+f3\right)\frac{dPrr}{Prr}+f8\cdot\frac{dPrH}{PrH}\\&=\frac{d\tau}{\tau}-\beta 2cw\cdot\frac{d\theta 1}{\theta 1}+\frac{\beta 1cw+\beta 1cw\beta 2-\beta 1((-1+\beta 1cw)\beta 1cw+(1+\beta 2cw)^2)}{2\beta 1\beta 1cw}\frac{d\theta 2}{\theta 2}\\&+\beta 2cw\cdot\left(\frac{d\theta 3}{\theta 3}+\frac{d\theta 4}{\theta 4}\right)-\frac{dPtr}{Ptr}+(1+\beta 2cw)\frac{dPtH}{PtH}+\frac{1+3\beta 1+\beta 2}{4\beta 1}\frac{dPir}{Pir}\\&-\frac{(1+\beta 1cw+\beta 2cw)^2}{4\beta 1cw}\frac{dPiH}{PiH}-\frac{1-\beta 1+\beta 2}{4\beta 1}\frac{dPrr}{Prr}+\frac{(1-\beta 1cw+\beta 2cw)^2}{4\beta 1cw}\frac{dPrH}{PrH}\end{aligned}$$

$$\frac{\Delta Q0}{Q0}=\sqrt{\left(\frac{\Delta\tau}{\tau}\right)^2+\beta 2cw^2\left(\frac{\Delta\theta 1}{\theta 1}\right)^2+\left(\frac{\beta 1cw+\beta 1cw\beta 2-\beta 1((-1+\beta 1cw)\beta 1cw+(1+\beta 2cw)^2)}{2\beta 1\beta 1cw}\right)^2\cdot\left(\frac{\Delta\theta 2}{\theta 2}\right)^2+\beta 2cw^2\cdot\left[\left(\frac{\Delta\theta 3}{\theta 3}\right)^2+\left(\frac{\Delta\theta 4}{\theta 4}\right)^2\right]+\left(\frac{\Delta Ptr}{Ptr}\right)^2+(1+\beta 2cw)^2\left(\frac{\Delta PtH}{PtH}\right)^2+\frac{(1+\beta 1cw+\beta 2cw)^4}{16\cdot\beta 1cw^2}\left(\frac{\Delta PiH}{PiH}\right)^2+\left(\frac{1+3\beta 1+\beta 2}{4\beta 1}\right)^2\left(\frac{\Delta Pir}{Pir}\right)^2+\frac{(1-\beta 1cw+\beta 2cw)^4}{16\cdot\beta 1cw^2}\left(\frac{\Delta PrH}{PrH}\right)^2+\left(\frac{1-\beta 1+\beta 2}{4\beta 1}\right)^2\left(\frac{\Delta Prr}{Prr}\right)^2}$$

The error on Eacc is derived as below. It gives the Eq. (13)

$$\begin{aligned}\frac{dEacc}{Eacc}&=\frac{1}{2}\frac{d\tau}{\tau}+\frac{1}{2}\frac{d\theta 1}{\theta 1}+\frac{(f1+f2+f3)}{2}\frac{d\theta 2}{\theta 2}+\frac{1}{2}\frac{d\theta 4}{\theta 4}-\frac{1}{2}\frac{dPtr}{Ptr}+\frac{1}{2}\frac{dPtH}{PtH}+\frac{1}{2}\left(\frac{f1}{2}+f2\right)\frac{dPir}{Pir}-\frac{1}{2}\left(\frac{f1}{2}+f3\right)\frac{dPrr}{Prr}\\&=\frac{1}{2}\left(\frac{d\tau}{\tau}+\frac{d\theta 1}{\theta 1}+\frac{1+\beta 1+\beta 2}{2\beta 1}\frac{d\theta 2}{\theta 2}+\frac{d\theta 4}{\theta 4}-\frac{dPtr}{Ptr}+\frac{dPtH}{PtH}+\frac{1+3\beta 1+\beta 2}{4\beta 1}\frac{dPir}{Pir}-\frac{1-\beta 1+\beta 2}{4\beta 1}\frac{dPrr}{Prr}\right)\end{aligned}$$

$$\frac{\Delta \mathrm{Eacc}}{\mathrm{Eacc}}=\frac{1}{2}\sqrt{\left(\frac{\Delta\tau}{\tau}\right)^2+\left(\frac{\Delta\theta1}{\theta1}\right)^2+\left(\frac{1+\beta1+\beta2}{2\beta1}\right)^2\left(\frac{\Delta\theta2}{\theta2}\right)^2+\left(\frac{\Delta\theta4}{\theta4}\right)^2+\left(\frac{\Delta Ptr}{Ptr}\right)^2+\left(\frac{\Delta PtH}{PtH}\right)^2+\left(\frac{1+3\beta1+\beta2}{4\beta1}\right)^2\left(\frac{\Delta Pir}{Pir}\right)^2+\left(\frac{1-\beta1+\beta2}{4\beta1}\right)^2\left(\frac{\Delta Prr}{Prr}\right)^2}$$

**Case 2: the cable loss on the RF driving cable is changed by heating effect, and the cable loss factor α1 is re-calibrated. But the Qe2 obtained before re-calibrating the cable is used to calculate Q0 and Eacc.**

In this case, there are two more independent variables PiD and PrD as defined in section 3. By replacing θ2 with $\theta5=\sqrt{\mathrm{PrD}/\mathrm{PiD}}$ when calculating Q0, and keeping θ2 unchanged inside Qe2, the uncertainty of Q0 could be calculated as in Eq. (C2).

$$\mathrm{Q0}=\mathrm{Qe2}\times\frac{\theta3\cdot\theta4\cdot\mathrm{PtH}}{\theta1(\theta5\mathrm{PiH}-\mathrm{PrH}/\theta5)-\theta3\theta4\mathrm{PtH}} \tag{C2}$$

Note the accuracy of Q0 is different from that in case 1, but Eacc is the same. By replacing θ2 with θ5, define f10, f11 and f12 as following:

$$\mathrm{f10}\triangleq\frac{\mathrm{Picw}}{\mathrm{Pccw}}=\frac{\theta1\cdot\theta5\cdot\mathrm{PiH}}{\theta1(\theta5\mathrm{PiH}-\mathrm{PrH}/\theta5)-\theta3\theta4\mathrm{PtH}}=\frac{(1+\beta1cw+\beta2cw)^2}{4\beta1cw}$$

$$\mathrm{f11}\triangleq\frac{\mathrm{Prcw}}{\mathrm{Pccw}}=\frac{\theta1/\theta5\cdot\mathrm{PrH}}{\theta1(\theta5\mathrm{PiH}-\mathrm{PrH}/\theta5)-\theta3\theta4\mathrm{PtH}}=\frac{(1-\beta1cw+\beta2cw)^2}{4\beta1cw}$$

$$\mathrm{f12}\triangleq\frac{\mathrm{Ptcw}}{\mathrm{Pccw}}=\frac{\theta3\cdot\theta4\cdot\mathrm{PtH}}{\theta1(\theta5\mathrm{PiH}-\mathrm{PrH}/\theta5)-\theta3\theta4\mathrm{PtH}}=\mathrm{\beta2cw}$$

The error on Q0 is then derived as below:

$$\begin{aligned}\frac{\mathrm{dQ0}}{\mathrm{Q0}}&=\frac{\mathrm{d\tau}}{\tau}+(1-f10+f11)\cdot\frac{d\theta1}{\theta1}+(f1+f2+f3)\frac{d\theta2}{\theta2}+f12\cdot\left(\frac{d\theta3}{\theta3}+\frac{d\theta4}{\theta4}\right)-(f10+f11)\frac{d\theta5}{\theta5}\\&-\frac{dPtr}{\mathrm{Ptr}}+(1+f12)\frac{dPtH}{PtH}+\left(\frac{f1}{2}+f2\right)\frac{dPir}{Pir}-f10\cdot\frac{dPiH}{PiH}-\left(\frac{f1}{2}+f3\right)\frac{dPrr}{Prr}+f11\cdot\frac{dPrH}{PrH}\\&=\frac{\mathrm{d\tau}}{\tau}-\beta2cw\cdot\frac{d\theta1}{\theta1}+\frac{1+\beta1+\beta2}{2\beta1}\frac{d\theta2}{\theta2}+\beta2cw\cdot\left(\frac{d\theta3}{\theta3}+\frac{d\theta4}{\theta4}\right)-\frac{\beta1cw^2+(1+\beta2cw)^2}{2\beta1cw}\frac{d\theta5}{\theta5}\\&-\frac{dPtr}{Ptr}+(1+\beta2cw)\frac{dPtH}{PtH}+\frac{1+3\beta1+\beta2}{4\beta1}\frac{dPir}{Pir}-\frac{(1+\beta1cw+\beta2cw)^2}{4\beta1cw}\frac{dPiH}{PiH}\\&-\frac{1-\beta1+\beta2}{4\beta1}\frac{dPrr}{Prr}+\frac{(1-\beta1cw+\beta2cw)^2}{4\beta1cw}\frac{dPrH}{PrH}\end{aligned}$$

$$\left.\frac{\Delta\mathrm{Q0}}{\mathrm{Q0}}\right|_{\mathrm{re-cal}}=\sqrt{\begin{aligned}&\left(\frac{\Delta\tau}{\tau}\right)^2+\beta2cw^2\left(\frac{\Delta\theta1}{\theta1}\right)^2+\left(\frac{1+\beta1+\beta2}{2\beta1}\right)^2\cdot\left(\frac{\Delta\theta2}{\theta2}\right)^2+\left(\frac{\beta1cw^2+(1+\beta2cw)^2}{2\beta1cw}\right)^2\cdot\left(\frac{\Delta\theta5}{\theta5}\right)^2\\&+\beta2cw^2\cdot\left[\left(\frac{\Delta\theta3}{\theta3}\right)^2+\left(\frac{\Delta\theta4}{\theta4}\right)^2\right]+\left(\frac{\Delta Ptr}{Ptr}\right)^2+(1+\beta2cw)^2\left(\frac{\Delta PtH}{PtH}\right)^2+\frac{(1+\beta1cw+\beta2cw)^4}{16\cdot\beta1cw^2}\left(\frac{\Delta PiH}{PiH}\right)^2\\&+\left(\frac{1+3\beta1+\beta2}{4\beta1}\right)^2\left(\frac{\Delta Pir}{Pir}\right)^2+\frac{(1-\beta1cw+\beta2cw)^4}{16\cdot\beta1cw^2}\left(\frac{\Delta PrH}{PrH}\right)^2+\left(\frac{1-\beta1+\beta2}{4\beta1}\right)^2\left(\frac{\Delta Prr}{Prr}\right)^2\end{aligned}}$$

**Appendix D: recall the wrong algorithm used in [1] to calculate error of Q0 and Eacc for decay measurement**

In the "Derivation of measurement errors – decay measurement" section in the Appendix A of [1], the error of Q0 and gradient is calculated as following:

$$\frac{\Delta\mathrm{Q0}}{\mathrm{Q0}}=\sqrt{\left(\frac{\Delta QL}{QL}\right)^2+\frac{\Delta\beta1^2+\Delta\beta2^2}{(1+\beta1+\beta2)^2}},\qquad\frac{\Delta\mathrm{Eacc}}{\mathrm{Eacc}}=\sqrt{\left(\frac{\Delta Q0}{Q0}\right)^2+\left(\frac{\Delta Pc}{Pc}\right)^2}$$

All the relative errors could be presented in the form of $\sqrt{\left(\frac{d\tau}{\tau}\right)^2+F(\beta1,\beta2)\times\left(\frac{\Delta P}{P}\right)^2}$, by adopting the same assumptions as described in the section 2 and 7, except that identical deviation of each power meter reading is assumed. Note all the calculations below follow the method in appendix A in [1]:

$$\frac{\Delta\mathrm{QL}}{\mathrm{QL}}=\frac{d\tau}{\tau}$$

$$\frac{\Delta\mathrm{Pc}}{\mathrm{Pc}}=\frac{1}{\mathrm{Pc}}\sqrt{\Delta Pi^2+\Delta Pr^2+\Delta Pt^2}=\sqrt{f4^2\left(\frac{\Delta Pi}{Pi}\right)^2+f5^2\left(\frac{\Delta Pr}{Pr}\right)^2+f6^2\left(\frac{\Delta Pt}{Pt}\right)^2}$$

Where f4, f5, and f6 are defined in Appendix B.

$$\frac{\Delta\beta2}{\beta2}=\sqrt{\left(\frac{\Delta Pt}{Pt}\right)^2+\left(\frac{\Delta Pc}{Pc}\right)^2}=\sqrt{1+f4^2+f5^2+f6^2}\frac{\Delta P}{P}$$

$$=\sqrt{\frac{\beta1^4+(1+\beta2)^4+2\beta1^2(7+\beta2(6+7\beta2))}{8\cdot\beta1^2}}\frac{\Delta P}{P}$$

$$\frac{\Delta\beta 1}{\beta 1} = \sqrt{\left(\frac{\Delta\beta^*}{\beta^*}\right)^2 + \left(\frac{\Delta\beta 2}{1+\beta 2}\right)^2}$$

$$\frac{\Delta\beta^*}{\beta^*} = \sqrt{\left(\frac{\Delta|\Gamma|}{1+|\Gamma|}\right)^2 + \left(\frac{\Delta|\Gamma|}{1-|\Gamma|}\right)^2} = \frac{|\Gamma|\sqrt{1+|\Gamma|^2}}{1-|\Gamma|^2}\frac{\Delta|\Gamma|}{|\Gamma|}$$

$$\frac{\Delta|\Gamma|}{|\Gamma|} = \frac{1}{2}\sqrt{\left(\frac{\Delta Pr}{Pr}\right)^2 + \left(\frac{\Delta Pi}{Pi}\right)^2} = \frac{\sqrt{2}}{2}\frac{\Delta P}{P}$$

$$|\Gamma| = \pm\frac{1-\beta 1+\beta 2}{1+\beta 1+\beta 2}$$

$$\therefore \frac{\Delta\beta^*}{\beta^*} = \pm\frac{(1-\beta 1^2+2\beta 2+\beta 2^2)\sqrt{1+\frac{(1-\beta 1+\beta 2)^2}{(1+\beta 1+\beta 2)^2}}}{4\beta 1+4\beta 1\beta 2}\frac{\Delta P}{P}$$

$$\frac{\Delta\beta 1}{\beta 1} = \sqrt{\frac{-2\beta 1^3(1+\beta 2)-2\beta 1(1+\beta 2)^3+\beta 1^4(1+\beta 2^2)+(1+\beta 2)^4(1+\beta 2^2) +2\beta 1^2(1+\beta 2(2+\beta 2(8+\beta 2(6+7\beta 2))))}{8\beta 1^2(1+\beta 2)^2}}\frac{\Delta P}{P}$$

Thus, the error of Q0 and Eacc are shown in Eq.

$$\frac{\Delta Q0}{Q0} = \sqrt{\left(\frac{d\tau}{\tau}\right)^2 + \frac{(\beta 1^2+(1+\beta 2)^2)(-2\beta 1^3(1+\beta 2)+\beta 2^2(1+\beta 2)^4+\beta 1^4(1+\beta 2^2) +\beta 1^2(1+\beta 2(2+\beta 2(15+2\beta 2(6+7\beta 2)))))}{8\beta 1^2(1+\beta 2)^2(1+\beta 1+\beta 2)^2}\left(\frac{\Delta P}{P}\right)^2} \tag{D1}$$

$$\frac{\Delta Eacc}{Eacc} = \sqrt{\left(\frac{d\tau}{\tau}\right)^2 + \frac{1}{16}\left[16\beta 2^2+\frac{(1-\beta 1+\beta 2)^4}{\beta 1^2}+\frac{(1+\beta 1+\beta 2)^4}{\beta 1^2} + \frac{2(\beta 1^2+(1+\beta 2)^2)(-2\beta 1^3(1+\beta 2)+\beta 2^2(1+\beta 2)^4+\beta 1^4(1+\beta 2^2) +\beta 1^2(1+\beta 2(2+\beta 2(15+2\beta 2(6+7\beta 2)))))}{\beta 1^2(1+\beta 2)^2(1+\beta 1+\beta 2)^2}\right]\times\left(\frac{\Delta P}{P}\right)^2} \tag{D2}$$